# Towards an accelerated decarbonization of chemical industry by electrolysis

**Magda H. Barecka[a], Joel W. Ager[b,c,d]**

a. Cambridge Centre for Advanced Research and Education in Singapore, CARES Ltd. 1 CREATE Way, CREATE Tower #05-05, 138602 Singapore; email: magda.barecka@cares.cam.ac.uk

b. Berkeley Education Alliance for Research in Singapore (BEARS), Ltd., 1 CREATE Way, 138602, Singapore

c. Department of Materials Science and Engineering, University of California at Berkeley, Berkeley, California 94720, USA

d. Materials Sciences Division, Lawrence Berkeley National Laboratory, Berkeley, California 94720, United States.

## Abstract

The transition towards carbon-neutral chemical production is challenging due to the fundamental reliance of the chemical sector on petrochemical feedstocks. Electrolysis-based manufacturing, powered by renewables, is a rapidly evolving technology that might be capable of drastically reducing $CO_2$ emissions from the chemical sector. However, will it be possible to scale up electrolysis systems to the extent necessary to entirely decarbonize all chemical plants? Applying a forward-looking scenario, this perspective estimates how much energy will be needed to power full-scale electrolysis-based chemical manufacturing by 2050. A significant gap is identified between the currently planned renewable energy expansion and the energy input necessary to electrify the chemical production: at minimum, the energy required for production of hydrogen and electrolysis of $CO_2$ corresponds to > 50% of all renewable energy that is planned to be available. To cover this gap, strategies enabling a meaningful reduction of the energy input to electrolysis are being discussed from the perspective of both a single electrolysis system and an integrated electro-plant. Several scale-up oriented research priorities are formulated to underpin timely development and commercial availability of described technologies, as well as to explore synergies and support further growth of the renewable energy sector, essential to realize described paradigm shift in chemical manufacturing.

## Keywords

## Introduction

Transition to carbon neutrality requires drastic changes to happen in an unprecedently short period of time[1]. Decarbonization of chemical manufacturing is crucial for achieving Net Zero 2050, as this sector is responsible for over 15% of all industrial $CO_2$ emissions[2]. The chemical industry is particularly challenging to decarbonize due to its fundamental reliance on the inputs of petrochemical feedstocks, used in two dominant ways. First, petrochemical resources (e.g., natural gas) are used as fuels for combustion, which is necessary to produce thermal energy to drive the chemical transformations. Secondly, feedstocks such as crude oil are being used as a starting material for the production of bulk chemicals, and the petrochemicals-derived carbon becomes embodied in the structure of final products. If the bulk chemicals are used to produce fuels, this carbon load will soon result in $CO_2$ emissions[3]; the same holds true for products with a short life-cycle such as e.g. plastics which are incinerated instead of being buried in the landfills. While these emissions do not necessarily happen within the physical boundary of the plant, they are a direct consequence of manufacturing strategies deployed at the production site (Fig. 1a).

To deeply decarbonize the chemical sector, we need to find a sustainable replacement for petrochemical resources, which at the same time delivers energy (through its chemical bonds), as well as carbon and hydrogen necessary to build complex products. In this context, emerging electrolysis technologies are particularly promising, as they use electric energy to drive chemical transformations; this energy can be sourced in a renewable manner. It also uses dilute, though naturally abundant materials such as $CO_2$ and water as feedstocks to produce chemicals (e.g. ethylene), hence is capable of being a sustainable replacement of petrochemical inputs to chemical manufacturing[4] (Fig 1b). In contrast to biomass-based methods used to produce chemicals, electrolysis does not require to sacrifice arable land, which is particularly scarce in some regions[5].

Given this great promise and the significant R&D interest in electrolysis[6], this perspective scrutinizes the potential of this technology to operate on scales necessary for the decarbonization of chemical industry, focusing on the necessary energy inputs for producing hydrogen and powering $CO_2$ electrolysis. The analysis presented in this paper highlights the extremely high, and so far likely underestimated, requirement for renewable energy to drive scaled-up electrolysis. In response to this challenge, diverse strategies allowing for up to two orders of magnitude reduction of the required renewable energy input are briefly introduced. The authors discuss as well how to maximize the reduction of $CO_2$ emissions across the entire chemical sector, operate electrolysis in a synergy with renewable energy production, and identify the features of electrolysis technology which need to be developed to facilitate further growth of renewables sector.

**Electrolysis types in chemical manufacturing**

Electrolysis technology is not completely unknown to chemical manufacturing as it has been used since 19$^{th}$ century for production of chlorine and sodium hydroxide (chloralkali process) and is widely deployed in electrometallurgy of, e.g., aluminium and lithium[7–9]. However, electrolysis-based methods did not successfully penetrate other manufactures[10,11]. Electrolysis technologies, that are currently investigated in the context of carbon neutrality, focus mostly on hydrogen production and $CO_2$ electroconversion to hydrocarbons. Hydrogen production is a relatively mature and scalable approach, available at Technology Readiness Level (TRL) of 9 (operational system) in Alkaline-type[12] or rapidly developing Polymer Electrolyte Membrane (PEM) electrolyzers[13–19]. Hydrogen generated by electrocatalytic methods can replace natural gas used as thermal energy vector and can eliminate the emissions arising from combustion of petrochemical sources. Importantly, the use of hydrogen as an alternative fuel is reported to require only minor retrofits to the existing furnaces, especially if hydrogen will be blended with some amount of natural gas. This holds a promise of reducing the carbon footprint of chemical manufactures with a limited retrofit cost and shut-down time[20]. Furthermore, renewably sourced hydrogen can be subsequently used as a co-feedstock in catalytic production of hydrocarbons from $CO_2$, with methanol synthesis being a well-understood and scalable example of such an approach[21].

Instead of deploying a two-step synthesis, renewably sourced hydrocarbons can be produced in a single step by a direct electrolysis of $CO_2$ to carbon monoxide, ethylene, methane, ethanol or propanol[4,22–25]. This technology allows for a simplified deployment of modular units that can yield bulk chemicals. Among different $CO_2$ electrolysis products, carbon monoxide/syngas[26] can be obtained from commercially available units (TRL 9) which use high-temperature solid oxide cells[27]; there are also several start-ups working towards the scale up of $CO_2$ to syngas in low-temperature stacks[28,29]. Production of ethylene[22,23,30], methane[31] and liquid fuels has been reported only on laboratory-scale so far, with a significant commercial interest into scale-up[6]. From a commercial perspective, ethylene is being a particularly promising electrolysis product due its high price in certain markets (Asia, European Union)[32]. It has been demonstrated that there exist a large number of applications where ethylene production by electrolysis could yield a remarkable economic benefit[33,34], with electricity prices below 0.045 $/kWh and performance metrics currently demonstrated in laboratory environment. In addition to that, there is a raising demand for green ethylene as a starting material for polymers synthesis, further used to manufacture carbon neutral products (e.g. apparel, fashion accessories), gaining popularity among climate-aware customers[35,36].

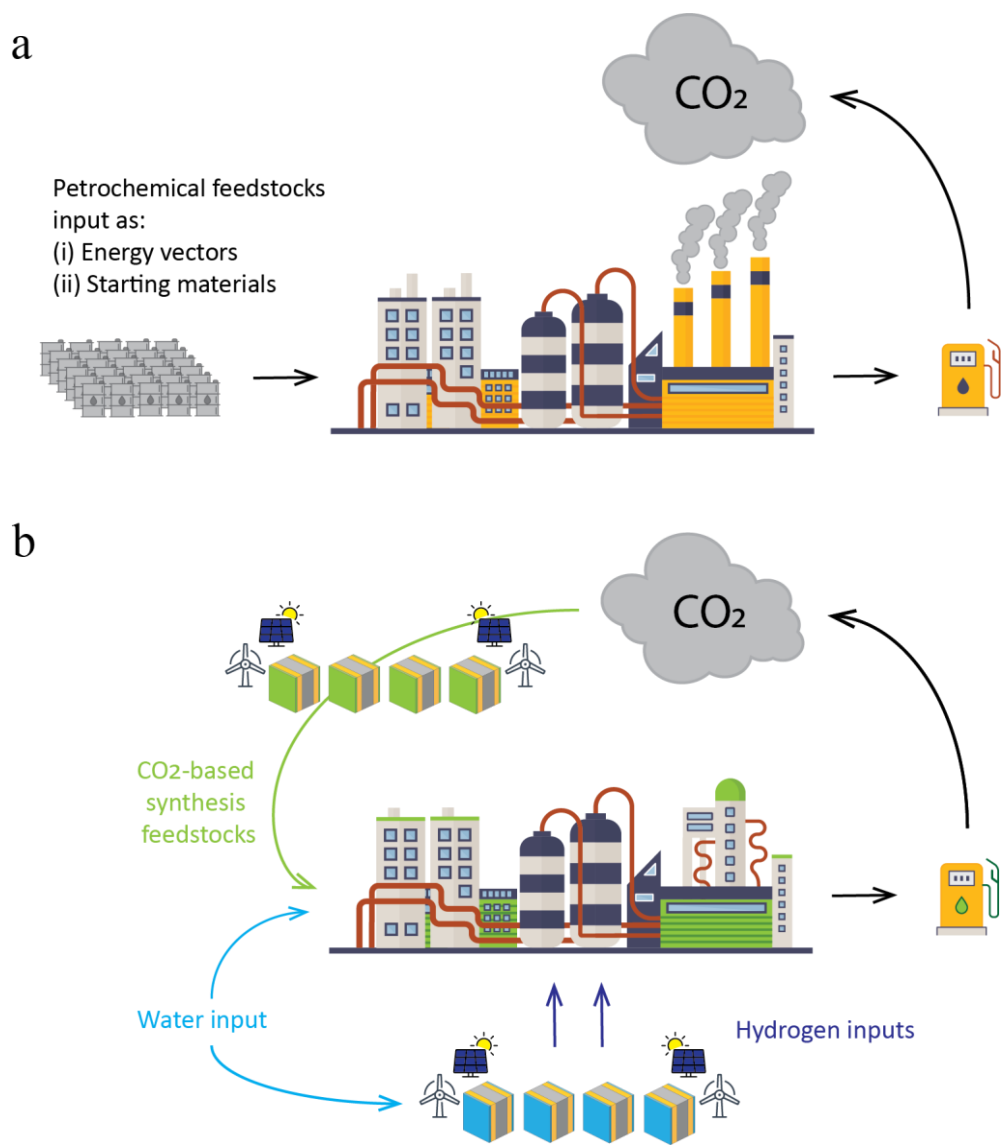

Fig. 1. $CO_2$ emissions from the chemical sector and pathways to their mitigation by electrolysis technologies: a) scheme of a typical chemical manufacturing plant, where petrochemical feedstocks are used to supply energy and starting materials, b) concept of a decarbonized manufacturing by means of electrolysis; hydrogen from water electrolysis is supplied as an energy vector, and feedstocks for chemical conversion processes are obtained by electrolysis of $CO_2$. To reduce life-cycle $CO_2$ emissions, all electrolyzers need to be powered by low-carbon energy.

**Energy required for large-scale electrolysis**

Taking into consideration how the chemical industry operates, both hydrogen and $CO_2$ electrolysis products will be necessary on large scales to provide carbon-neutral energy sources and feedstock materials. With a significant progress in the development of all electrolysis processes described above, it is timely to question if we are planning for a sufficient expansion of renewable energy to power these processes on the scales necessary by 2050. Deploying a simple assessment and assuming that all natural gas input to chemical manufactures[37] shall be replaced by hydrogen, we quantified the required hydrogen input (as it has different energy density than natural gas), and consequently the necessary electric energy to generate this amount of hydrogen (based on forward-looking electrolyzers efficiency of 85%). The resulting energy requirement is ~$5 \cdot 10^{19}$ J/year and accounts for 44% of the projected renewable energy generation in 2050 as reported by the International Renewable Energy Agency[38]. Other agencies[39] report planned expansion of renewable energy generation in the same order of magnitude.

A similar analysis was deployed to assess the required energy for electrocatalytic production of ethylene, on the scale necessary to replace all petrochemically-derived ethylene used as a starting material for chemical synthesis (185 Mt/year[40]). With hypothetical electrolyzer efficiency of 85%, ~$1 \cdot 10^{19}$ J/year of energy would be needed, being 10% of what is planned to be available from renewable sources in 2050. Importantly, this assessment is done based on current ethylene (and natural gas) consumption; with growing population and needs across food, health and personal care, the demand for raw materials will also rise.

In total, the presented analysis points out that a minimum energy consumption for production of hydrogen and electrolysis of $CO_2$ would consume >50% of all renewable energy that is planned to be available, leaving less than 50% capacity to power electrified transportation, commercial/public services, residential, food production, data centres and other manufacturing sectors. Therefore, the current plans on renewable energy generation (on global level) might be not sufficient to allow for deep decarbonization of the chemical industry by means of electrolysis and will consequently be a rate-limiting factor in decarbonization efforts. Hence, we need to either plan the expansion of the renewable electricity production much more boldly or (and) need drastic reductions in terms of the energy input to the electrolysis units. This perspective proposes multiple emerging research areas which can yield scalable technologies that respond to this challenge and allow for production of chemicals and fuels with a minimized input of renewable energy.

**Pathways to energy input reduction: hydrogen**

Drastic reduction of the energy input to hydrogen production requires looking beyond the currently deployed chemistry. One of the ways to reduce this input is to study other reactions than water splitting, which is thermodynamically bounded to min. 1.23 V energy input at standard temperature and pressure. The energy-intensive anodic oxygen evolution reaction can be replaced by electro-oxidation, that instead of pure water uses liquid biomass derivates, alcohols or amines[41] (Fig. 2a). As a result, the thermodynamic cell voltage requirement can be drastically reduced[42,43] (e.g. to 0.3 V). This drastic reduction of energy requirement for hydrogen production can be a paradigm shift in large scale deployment of electrocatalytic systems and at the same time allows to generate value-added chemicals on the anode side (instead of oxygen stream, or $CO_2$ which would be the case if the biomass feedstock was completely oxidized).

Consequently, it is important to understand how to sustainably source large quantities of the molecules which could be a convenient anodic feedstock. Coupling hydrogen production with industrial wastewater treatment is promising in this regard. This concept was proposed by Qiu et al [44] for the case of pulping industry, which yields a wide range of carbohydrate alkaline degradation products (CHADs), and such simulated waste stream was demonstrated to be a functional feed for hydrogen production. Sustainably sourced alcohols are also a promising anodic input[45]; this pathway can be particularly promising if ethanol waste streams could be sourced from waste streams from biomanufacturing, as these industries are likely to scale-up in the upcoming decades. Wang et al.[46] have further explored this concept by proposing anodic oxidation reactions that also yield hydrogen product, hence the overall energy requirement is not only decreased by a more thermodynamically favourable anodic transformation, but also because of increased hydrogen output from the same unit (Fig. 2b). The limitations in the scaling of the approach that involves the use of alternative feedstocks on the anode side are described in "Limitations of the study" section.

Another possibility for reduction of the electrical energy input relates to inclusion of enzymes and development of combined electrochemistry-enzyme systems (Fig. 2c). Enzymes has been extensively studied for their excellent catalytic properties and as a sustainable replacement to catalysts that would need to be otherwise mined, like e.g. noble metals[47]. In the field of hydrogen production, numerous enzyme-electroreduction systems were investigated for their potential to deliver improvement in energy efficiency[48–50]. Hardt et al.[51] reported an example of such enzymatic system (hydrogenase embedded in a hydrogel), allowing to produce hydrogen with only 12 mV overpotential.

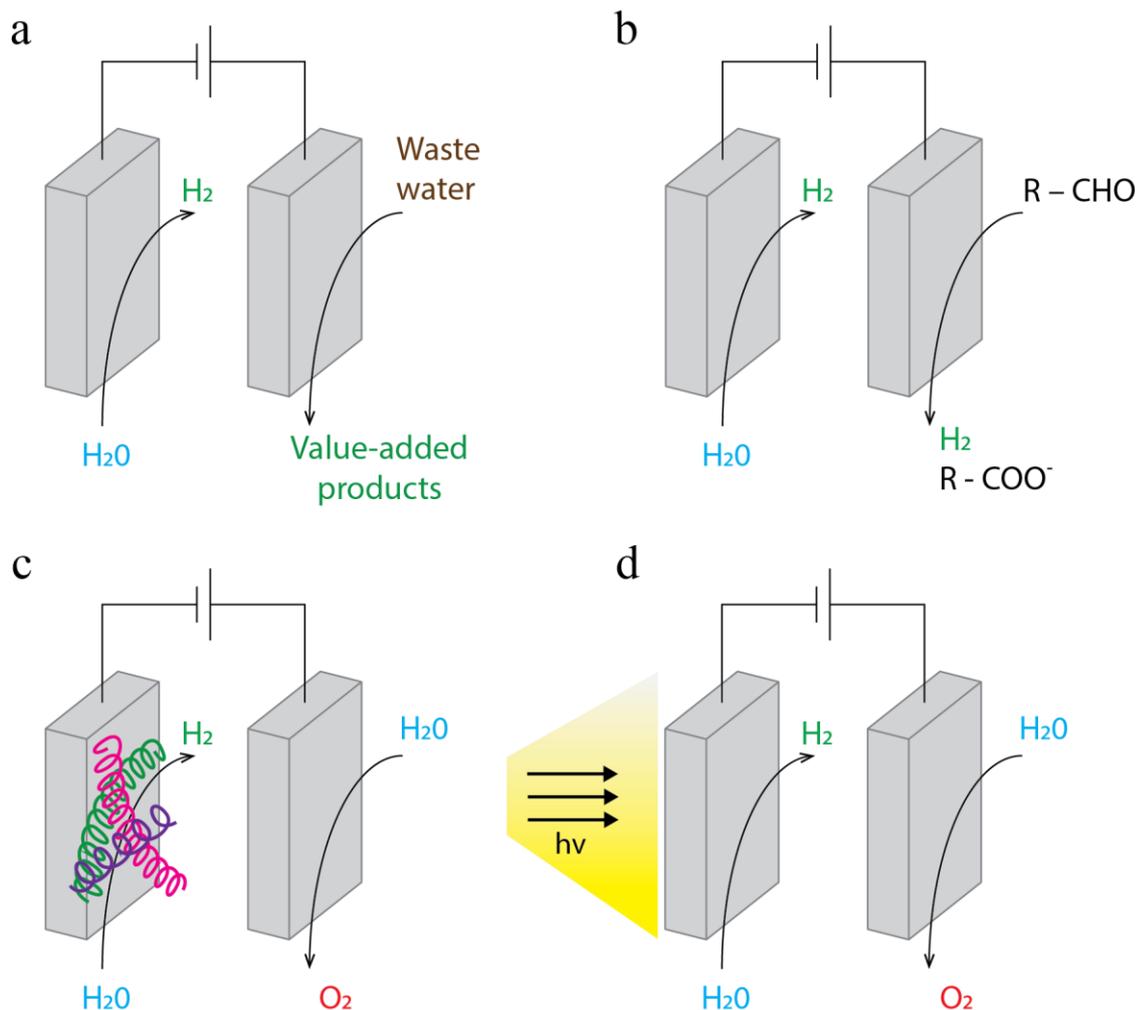

Fig.2. Conceptual representation of exemplary strategies allowing for a reduction of energy input to electrolysis processes, exemplified for hydrogen production by means of: a) combination with anodic oxidation, b) co-production of hydrogen on the anode side, c) incorporation of enzymes, d) inclusion of solar radiation energy inputs.

Furthermore, system energy requirements can be reduced by inclusion of additional energy sources beyond the externally sourced electricity. This approach includes the use of bacteria, that could produce electricity used to drive the electrochemical reaction (like in microbial electrolysis cells); however, bacteria emit $CO_2$ that will need to be captured and stored[52]. Alternatively, bacteria could be deployed to directly produce hydrogen, which could be further intensified by combination with electrocatalytic methods. Vasiliadou et al.[53] describes such an approach, which combines the capacity of purple phototrophic bacteria to produce hydrogen with electrochemical reaction on a working electrode of graphite.

Another pathway to reduce external inputs of electricity to electrocatalytic systems uses energy from sunlight by means of photocatalytic systems, which has been demonstrated on a scale of up to 100 m$^2$ of electrode surface[54] (Fig. 2d). However, these devices at present are much less productive in terms of hydrogen output per surface area as compared to solely electric-energy powered electrolyzers (e.g. PEM), hence extremely large reactors will be required to deliver the same hydrogen throughout. Resulting high investment costs, along with extended requirement for space and materials necessary to build electrodes highlights the need to focus on increasing the productivity and seek for widely available, or easily recyclable electrodes materials[55].

**Pathways to energy input reduction: $CO_2$ electrolysis**

All pathways towards the reduction of energy input described above are applicable not only to hydrogen production, but as well to $CO_2$ electrolysis[56]. Na et al. has scrutinized the possibility of replacement of water feed on the anode side by different organic waste streams leading to co-production of useful chemicals on the anode side, and most importantly, to a drastic reduction in the energy input. Their analysis discusses the opportunities to produce 13 different chemicals on the cathode side and 20 on the anode side. Remarkably, combination of $CO_2$ electrolysis to ethylene on the cathode side with glycerol oxidation to formic acid on the anode side leads to reduction of the full cell voltage from 1.15 V to 0.06 V[57] (under hypothetical 100% efficiency). This two-orders of magnitude reduction in energy requirement opens pathway to electrolysis deployment on unprecedented levels.

Inclusion of enzymes[58], bacteria[59], and additional energy sources[60] has also been proposed for $CO_2$ electrolysis application. Interestingly, there are also insights into combination of some or all of these functionalities into one device, inspired by photosynthesis process. Possibilities of enzymatic electroreduction were studied to deliver carbon monoxide, formic acid and methanol[61]; promising findings in terms of the energy input reduction, excellent selectivity and stability were reported for carbon monoxide production by Carbon Monoxide Dehydrogenase from Moorella thermoacetica[62]. The energy barrier could be further reduced by including the direct energy input from solar radiation in photo-bio electrocatalytic devices[63].

**Eliminating several sources of emissions with the same energy input**

Another approach to support the deployment of electrolysis is to focus on mitigating several sources of $CO_2$ emissions with one electrocatalytic device, increasing the amount of avoided $CO_2$ emissions per unit of energy input to the processes. This could be achieved by a direct, one-step $CO_2$ electrolysis to complex molecules, being these the final output of the chemical manufacture (e.g. ethylene glycol). As a result, the electrolysis enables to bypass the entire chemical plant, avoiding $CO_2$ emissions from the use of petrochemical resources as energy vectors and production feedstocks. Hence, there is no need to separately invest energy to produce hydrogen and sustainably-sourced hydrocarbon feedstocks; instead, a single energy input, along with $CO_2$ and water, will suffice for the operation of the entire electro-plant. So far, this approach has been explored on laboratory scale production of ethylene oxide, propylene oxide[64] and ethylene glycol[65]. More research is necessary to extend this portfolio, and importantly, these insights should be carried out in devices applicable for large-scale electrolysis (gas diffusion electrodes assembly) instead of H-cell type electrolyzers preferred for the convenience of laboratory studies[66]. Ideally, development of electrocatalytic routes towards more complex chemicals should be combined with any approach that reduces energy requirements: enzymes use or alternative anode-side reactions.

**Synergistic support for renewables expansion**

Direct production of complex chemicals from $CO_2$ can also support further investment into new renewable electricity projects by providing a strategy to address the current limitations towards the expansion of renewables: the high cost of energy transmission and battery storage[67]. First, being able to produce chemical products directly next to the power plant eliminates the energy transmission cost;

however, targeted products should be either used locally, either easy and safe to transport. Secondly, the ability to operate electrochemical processes in an intermittent manner can also remove the need for costly and material-intense battery storage, as the produced energy can be directly consumed on site only when it is generated. Low-temperature electrolysis, itself, is perfectly suited to operate only upon the availability of renewable energy, with a start-up and shut down times in a range of seconds to minutes. However, to deploy this concept, it is necessary also to run intermittently further transformations of the electrolysis product. This is particularly challenging to achieve if one envisions e.g. further conversion of hydrogen or syngas by means of high-temperature catalytic methods which are associated with extended start-up and shut-down time. Thus, the ability to directly produce complex chemicals solves this problem, as no further chemical transformations are necessary. Availability of such modular electro-plants could foster the investment into new renewable energy projects, well beyond what is currently planned to be executed by 2050.

**Emerging research priorities**

Based on the stated above, there exist a wide range of options that could allow for a deep decarbonization of the chemical industry by the electrochemical technology. Notably, all of the described approaches have been verified only on the laboratory scale, and their further development needs to happen in a very short period of time. Therefore, it is timely to review the research goals and formulate strategies that will support an accelerated scale-up.

First, despite of the fact that the most mature electrolysis technologies (water splitting and $CO_2$ electrolysis with water feed at the anode side) are much more energy intensive than the alternatives discussed in this perspective, we should not refrain from scaling up these higher TRL options and deploying for a commercial production of bulk chemicals. It has been demonstrated that if well connected to the existing value chain, these technologies can be economically viable with current renewable electricity prices, even with no $CO_2$ taxation in place[33]. Use of the electrolysis on a large scale will yield unique knowledge and experience, necessary to develop other systems. Most importantly, scaling up these technologies now means that by the time that alternative electrolysis approaches will gain better understanding, large electrolyzers will be already commissioned and integrated within chemical plants. Therefore, at that point in future, it will be simpler and faster to retrofit existing "classical" electrodes by more energy efficient next generation of materials, or to deploy alternative anodic feeds and energy co-sources (Fig. 3). This parallel electrolysis scale-up, along with the expansion of the renewable electricity could allow to meet the goal of carbon neutral production by 2050.

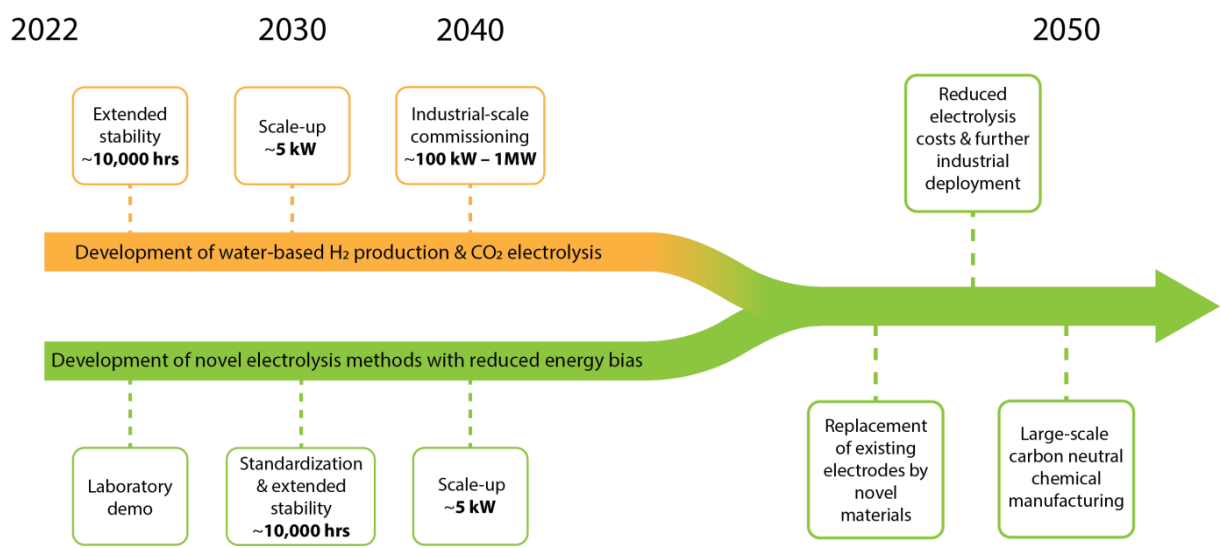

Fig. 3. Proposed phases for electrolysis systems scale-up: in parallel, we need to scale and commission the most mature systems based on water electrolysis and perform further research on energy-efficient alternatives described in this perspective. This will allow in future to quickly adopt already deployed electrolyzers once better alternatives are developed and save time as opposed to a classical, consecutive scale-up approach. Proposed targets related to the stability and scale-up were inspired by the reports for commercially available, small-scale hydrogen electrolyzers[68], and techno-economic analysis for hydrogen production published by the National Renewable Energy Laboratory[69].

Equally important, we need a coordinated development of the described technologies. Building up a common understanding of the experimental procedures for electrolysis characterization will help to quantify the status of development of each electrolysis variant. To this end, standards and protocols that detail the testing environment (e. g. reactor architecture, conditioning, operating conditions such as current density, feed composition etc.) are required. While protocols for standardized testing of commercial water electrolyzers have been proposed recently by the Joint Research Centre of the EU[70] and the National Renewable Energy Laboratory in the US[71], further discussion is necessary to understand how to make these protocols deployable for the assessment of less mature technologies, available at a laboratory scale. There is also an ongoing discussion in $CO_2$ electrolysis field regarding how to ensure a precise product quantification through the test rig design[72], how to test catalysts under the conditions relevant for industrial application[66], and how much the performance of the catalyst can be hindered by the use of suboptimal testing environment[73]. These examples highlight the importance of standardization in the electrolysis field, and it will be crucial to define a structure of protocol covering all these aspects in one testing procedure. Deploying such a protocol for every new material will consequently facilitate transparent comparison between electrolysis variants and understanding of related limitations. So far, rigorous protocols were proposed in $CO_2$ electrolysis field only for techno-economic analysis[74], and their availability is likely to support the acquisition of funding for further development.

We also need to anticipate scale-up challenges already within the early laboratory tests and focus the research on solving the critical problems towards development of functional systems. As such, we should test the operation of devices with industrial streams, that might include some minor, particularly challenging impurities. Running early-development tests with e.g. waste water feeds for hydrogen production, or $CO_2$ from industrial source points for $CO_2$ electrolysis would help to verify potential catalyst poisoning effects. While sourcing $CO_2$ from an industrial emitter is feasible to be done by using bottled gas, direct use of waste water is more restricted in research facilities. Therefore, developing test units that could be distributed to e.g. waste water treatment plants for testing with "real" streams should become a part of an early development procedure. Detailed insights into the durability of electrodes as well as sourcing and recycling of materials for their construction will be also pivotal to achieve ambitious scale-up goals. So far, the stability of electrodes is usually being verified on the time scale of hours to days, what needs to be increased by a factor of 100-1000 to open commercially viable applications. Frequently, stability studies are hindered by the lack of the infrastructure suited for the long-term testing, that will allow for a safe, unsupervised execution of experiments involving production gases such as hydrogen or carbon monoxide. Thus, development of well-sealed, thoroughly monitored and automated set-ups for electrolysis tests could support scrutinizing electrodes durability, which is particularly challenging in $CO_2$ electroreduction field.

While developing new electrolysis systems, it is also crucial to balance the academic pursuit towards excellence with a practical, hands-on approach. Though an idealistic 100% Faradaic efficiency (FE) towards a single product of electroreduction is an exciting research goal, it might be more insightful to appreciate what is the minimum FE necessary for a commercially attractive process and prioritize the work towards functional though less selective processes.

To further understand what is the minimum selectivity or conversion of electrolysis process that will yield a commercially attractive application, we need to question how electrolysis will operate as a part of the established value chain. Especially while thinking of electrolysis coupled with anodic oxidation, a connection to specific waste stream will be required, and electrolysis will deliver at least two different products. It is crucial to consider the availability of these waste streams, that varies locally, and will change with any transformation of the chemical industry, as well the possibilities of the use of anodic products. Therefore, it is reasoned to work towards a wide portfolio of reactions that can be deployed to produce hydrogen, and will allow both, uptake various, locally available waste sources and deliver anodic side by-products which will be easy to separate and useful in the given environment. Overall, from the perspective of large-scale electrolysis adoption, it might be more impactful to have access to less optimized, but a wide range of electrochemical reactions functional for e.g. hydrogen production, rather than having a sole access to thoroughly optimized water electrolyzers, which will in any case require vast energy input.

Last, we need to support the exchange of ideas and learnings between hydrogen production and $CO_2$ electrolysis fields. There are fundamental similarities between these two systems, however the research frequently happens separately. Deployment of unified testing protocols and standards might also help to intensify this collaboration. It is important to appreciate that ultimately, we will likely need a portfolio of water and $CO_2$ electrolysis options, rather than rely on one well developed technology. Unavailability of these diverse and scalable options that can suit production needs at a particular manufacture increases the long-term risk of not meeting the carbon neutrality goal.

**Limitations of the study**

The present analysis focuses only on the necessary energy inputs to electrolyzers and does not consider other energy needs across the life cycle, such as e.g. the cost of $CO_2$ capture and transport, inevitable for the electrolysis reactors that operate with concentrated $CO_2$ streams[75]. Large-scale deployment of this technology would thus require a capacity to pre-concentrate biogenic $CO_2$, which though abundant, has a much lower concentration that the industrial source points (with an exception of some fermentation-based processes[76]). Consequently, the total energy requirement of future electro-manufactures will be even higher, which further emphasises the need to focus on the energy efficiency aspects. We anticipate that though significant improvements have been reported in terms of the cost and scalability of direct air/biogenic sources capture[77], ensuring the availability of concentrated $CO_2$ will be another challenge that the electrolysis community needs to address; a country-wide analysis of biogenic $CO_2$ sourcing, related challenges and research priorities have been recently published by Badgett et al[78].

Responding to the growing renewable energy needs will also come at a significant capital cost, which is not quantified in our analysis. The investment in the manufacturing of solar panels (or wind mills), construction and maintenance of the new infrastructure for power distribution will strongly influence the feasibility of deployment of large scale electro-manufacturing envisioned here.

Pursuing some of the described strategies for the minimized energy input to hydrogen production, such as the use of waste streams as anode feedstocks, will also require re-inventing the way we optimize production processes on scale. Instead of working with a single set of feeds and products, envisioned here electro-manufactures will need to flexibly uptake waste streams with varying availabilities, as well as to be able to adjust their production strategies to the fluctuating market demand for the anode side by-products. Thus, in parallel to increasing the TRL of electrolysis technologies, we also need to develop new process design methods that will respond to the rising complexity of the supply chain. Such methods need to be capable of optimizing production routes through various feedstocks, both in terms of cost and the sustainability of the overall process. Availability of such planning tools on regional, or even larger scales, could substantially contribute to the use of electrolysis as tool for a fully circular, decarbonized manufacturing.

## Declaration of Competing interests

The authors co-filed patent applications US 62/987,369, US 63/036,477 and US 63/213,936 for novel electroreduction-based processes, e.g. for $CO_2$ recycling in ethylene-based plants. There are no other conflicts of interest to declare.

## Acknowledgements

The authors acknowledges the support by the National Research Foundation (NRF), Prime Minister's Office, Singapore under its Campus for Research Excellence and Technological Enterprise (CREATE) Programme through the eCO2EP project, operated by the Cambridge Centre for Advanced Research and Education in Singapore (CARES) and the Berkeley Educational Alliance for Research in Singapore (BEARS). The contribution of Andres J. Sanz Guillen to discussions, proofreading and the enrichment of the visual content is gratefully acknowledged.

## References

1  J. H. Wesseling, S. Lechtenböhmer, M. Åhman, L. J. Nilsson, E. Worrell and L. Coenen, The transition of energy intensive processing industries towards deep decarbonization: Characteristics and implications for future research, *Renewable and Sustainable Energy Reviews*, 2017, **79**, 1303–1313.

2  P. Levi, Technology Roadmap The global iron and steel sector. International Energy Agency., 2019.

3  P. Bains, P. Psarras and J. Wilcox, CO 2 capture from the industry sector, *Progress in Energy and Combustion Science*, 2017, **63**, 146–172.

4   J. W. Ager and A. A. Lapkin, Chemical storage of renewable energy, *Science (New York, N.Y.)*, 2018, **360**, 707–708.

5   V. Piemonte and F. Gironi, Land-use change emissions: How green are the bioplastics?, *Environ. Prog. Sustainable Energy*, 2011, **30**, 685–691.

6   S. Garg, M. Li, A. Z. Weber, L. Ge, L. Li, V. Rudolph, G. Wang and T. E. Rufford, Advances and challenges in electrochemical CO 2 reduction processes: an engineering and design perspective looking beyond new catalyst materials, *J. Mater. Chem. A*, 2020, **8**, 1511–1544.

7   G. Botte, Electrochemical Manufacturing in the Chemical Industry, *Interface magazine*, 2014, **23**, 49–55.

8   D. Pletcher, *Industrial electrochemistry*, Chapman and Hall, London, New York, 1984, 1982.

9   I. Moussallem, J. Jörissen, U. Kunz, S. Pinnow and T. Turek, Chlor-alkali electrolysis with oxygen depolarized cathodes: history, present status and future prospects, *J Appl Electrochem*, 2008, **38**, 1177–1194.

10  B. A. Frontana-Uribe, R. D. Little, J. G. Ibanez, A. Palma and R. Vasquez-Medrano, Organic electrosynthesis: a promising green methodology in organic chemistry, *Green Chem.*, 2010, **12**, 2099.

11  H. Lund, A Century of Organic Electrochemistry, *J Appl Electrochem*, 2002, **149**, S21.

12  Jensen, J.O., Bandur, V., Bjerrum, N.J., Jensen, S.H., Ebbesen, S., Mogensen, M., Tophøj, N., and Yde, L., Pre-investigation of water electrolysis. Report. PSO-F&U 2006-1-6287, 2006.

13  S. Shiva Kumar and V. Himabindu, Hydrogen production by PEM water electrolysis – A review, *Materials Science for Energy Technologies*, 2019, **2**, 442–454.

14  A. Ursua, L. M. Gandia and P. Sanchis, Hydrogen Production From Water Electrolysis: Current Status and Future Trends, *Proc. IEEE*, 2012, **100**, 410–426.

15  K. Zeng and D. Zhang, Recent progress in alkaline water electrolysis for hydrogen production and applications, *Progress in Energy and Combustion Science*, 2010, **36**, 307–326.

16  W. KREUTER, Electrolysis: The important energy transformer in a world of sustainable energy, *International Journal of Hydrogen Energy*, 1998, **23**, 661–666.

17  T. Smolinka, *"Water Electrolysis" in Encyclopedia of electrochemical power sources. Garche, Jürgen (Eds.)*, Elsevier, Amsterdam, 2009.

18  R. LEROY, Industrial water electrolysis: Present and future, *International Journal of Hydrogen Energy*, 1983, **8**, 401–417.

19  M. Paidar, V. Fateev and K. Bouzek, Membrane electrolysis—History, current status and perspective, *Electrochimica Acta*, 2016, **209**, 737–756.

20  M. Mayrhofer, M. Koller, P. Seemann, R. Prieler and C. Hochenauer, Assessment of natural gas/hydrogen blends as an alternative fuel for industrial heat treatment furnaces, *International Journal of Hydrogen Energy*, 2021, **46**, 21672–21686.

21  R. Guil-López, N. Mota, J. Llorente, E. Millán, B. Pawelec, J. L. G. Fierro and R. M. Navarro, Methanol Synthesis from CO2: A Review of the Latest Developments in Heterogeneous Catalysis, *Materials (Basel, Switzerland)*, 2019, **12**. DOI: 10.3390/ma12233902.

22  F. P. García de Arquer, C.-T. Dinh, A. Ozden, J. Wicks, C. McCallum, A. R. Kirmani, D.-H. Nam, C. Gabardo, A. Seifitokaldani, X. Wang, Y. C. Li, F. Li, J. Edwards, L. J. Richter, S. J. Thorpe, D. Sinton and E. H. Sargent, CO2 electrolysis to multicarbon products at activities greater than 1 A cm-2, *Science (New York, N.Y.)*, 2020, **367**, 661–666.

23  C.-T. Dinh, T. Burdyny, M. G. Kibria, A. Seifitokaldani, C. M. Gabardo, F. P. García de Arquer, A. Kiani, J. P. Edwards, P. de Luna, O. S. Bushuyev, C. Zou, R. Quintero-Bermudez, Y. Pang, D. Sinton and E. H. Sargent, CO2 electroreduction to ethylene via hydroxide-mediated copper catalysis at an abrupt interface, *Science (New York, N.Y.)*, 2018, **360**, 783–787.

24  J. E. Huang, F. Li, A. Ozden, A. Sedighian Rasouli, F. P. García de Arquer, S. Liu, S. Zhang, M. Luo, X. Wang, Y. Lum, Y. Xu, K. Bertens, R. K. Miao, C.-T. Dinh, D. Sinton and E. H. Sargent, CO2 electrolysis to multicarbon products in strong acid, *Science (New York, N.Y.)*, 2021, **372**, 1074–1078.

25  M. Li, M. N. Idros, Y. Wu, T. Burdyny, S. Garg, X. S. Zhao, G. Wang and T. E. Rufford, The role of electrode wettability in electrochemical reduction of carbon dioxide, *J. Mater. Chem. A*, 2021, **9**, 19369–19409.

26  W. Guo, J. Bi, Q. Zhu, J. Ma, G. Yang, H. Wu, X. Sun and B. Han, Highly Selective CO 2 Electroreduction to CO on Cu–Co Bimetallic Catalysts, *ACS Sustainable Chem. Eng.*, 2020, **8**, 12561–12567.

27  C. Mittal, C. Hadsbjerg and P. Blennow, Small-scale CO from CO2 using electrolysis, *Chemical Engineering World*, 2017, 44–46.

28  Dioxycle, Company website. https://dioxycle.com, 2022.

29  Twelve, Company website. https://www.twelve.co/, 2022.

30  L. Zaza, K. Rossi and R. Buonsanti, Well-Defined Copper-Based Nanocatalysts for Selective Electrochemical Reduction of CO 2 to C 2 Products, *ACS Energy Lett.*, 2022, **7**, 1284–1291.

31  A. Sedighian Rasouli, X. Wang, J. Wicks, G. Lee, T. Peng, F. Li, C. McCallum, C.-T. Dinh, A. H. Ip, D. Sinton and E. H. Sargent, CO 2 Electroreduction to Methane at Production Rates Exceeding 100 mA/cm 2, *ACS Sustainable Chem. Eng.*, 2020, **8**, 14668–14673.

32  ICIS, 2021 Global Market Outlook – Chemicals. Independent Commodity Intelligence Services report., 2020.

33  M. H. Barecka, J. W. Ager and A. A. Lapkin, Carbon neutral manufacturing via on-site CO2 recycling, *iScience*, 2021, **24**, 102514.

34  M. H. Barecka, J. W. Ager and A. A. Lapkin, Economically viable CO 2 electroreduction embedded within ethylene oxide manufacturing, *Energy Environ. Sci.*, 2021, **14**, 1530–1543.

35  S. J. Davis, N. S. Lewis, M. Shaner, S. Aggarwal, D. Arent, I. L. Azevedo, S. M. Benson, T. Bradley, J. Brouwer, Y.-M. Chiang, C. T. M. Clack, A. Cohen, S. Doig, J. Edmonds, P. Fennell, C. B. Field, B. Hannegan, B.-M. Hodge, M. I. Hoffert, E. Ingersoll, P. Jaramillo, K. S. Lackner, K. J. Mach, M. Mastrandrea, J. Ogden, P. F. Peterson, D. L. Sanchez, D. Sperling, J. Stagner, J. E. Trancik, C.-J. Yang and K. Caldeira, Net-zero emissions energy systems, *Science (New York, N.Y.)*, 2018, **360**. DOI: 10.1126/science.aas9793.

36  A. Peters, The lenses in these sunglasses are made from captured CO2. https://www.fastcompany.com/90678020/the-lenses-in-these-sunglasses-are-made-from-captured-co2, *Fastcompany*, 2021.

37 IEA, Gas. Report by the International Energy Agency. https://www.iea.org/fuels-and-technologies/gas, 2022.

38 International Renewable Energy Agency, Global energy transformation. Roadmap to 2050., 2018.

39 International Energy Agency, Net Zero by 2050. A Roadmap for the Global Energy Sector., 2021.

40 L. Fernandez, Global demand of ethylene 2017-2022, *Statista.com*, 2021.

41 M. Xiang, N. Wang, Z. Xu, H. Zhang and Z. Yan, Accelerating Hydrogen Evolution by Anodic Electrosynthesis of Value-Added Chemicals in Water over Non-Precious Metal Electrocatalysts, *ChemPlusChem*, 2021, **86**, 1307–1315.

42 Y. Xu and B. Zhang, Recent Advances in Electrochemical Hydrogen Production from Water Assisted by Alternative Oxidation Reactions, *ChemElectroChem*, 2019, **6**, 3214–3226.

43 Y. X. Chen, A. Lavacchi, H. A. Miller, M. Bevilacqua, J. Filippi, M. Innocenti, A. Marchionni, W. Oberhauser, L. Wang and F. Vizza, Nanotechnology makes biomass electrolysis more energy efficient than water electrolysis, *Nature communications*, 2014, **5**, 4036.

44 Z. Qiu, D. Martín-Yerga, P. A. Lindén, G. Henriksson and A. Cornell, Green hydrogen production via electrochemical conversion of components from alkaline carbohydrate degradation, *International Journal of Hydrogen Energy*, 2021, **22**, 4115.

45 H. Ju, S. Giddey, S. P. Badwal and R. J. Mulder, Electro-catalytic conversion of ethanol in solid electrolyte cells for distributed hydrogen generation, *Electrochimica Acta*, 2016, **212**, 744–757.

46 T. Wang, L. Tao, X. Zhu, C. Chen, W. Chen, S. Du, Y. Zhou, B. Zhou, D. Wang, C. Xie, P. Long, W. Li, Y. Wang, R. Chen, Y. Zou, X.-Z. Fu, Y. Li, X. Duan and S. Wang, Combined anodic and cathodic hydrogen production from aldehyde oxidation and hydrogen evolution reaction, *Nature catalysis*, 2021, **488**, 294.

47 N. Adam, S. Schlicht, Y. Han, M. Bechelany, J. Bachmann and M. Perner, Metagenomics Meets Electrochemistry: Utilizing the Huge Catalytic Potential From the Uncultured Microbial Majority for Energy-Storage, *Frontiers in bioengineering and biotechnology*, 2020, **8**, 567.

48 P. A. Ash and K. A. Vincent, in *Encyclopedia of Interfacial Chemistry*, Elsevier, 2018, pp. 590–595.

49 M. H. Prechtl and U.-P. Apfel, Toward electrocatalytic chemoenzymatic hydrogen evolution and beyond, *Cell Reports Physical Science*, 2021, **2**, 100626.

50 J. C. Ruth, F. M. Schwarz, V. Müller and A. M. Spormann, Enzymatic Hydrogen Electrosynthesis at Enhanced Current Density Using a Redox Polymer, *Catalysts*, 2021, **11**, 1197.

51 S. Hardt, S. Stapf, D. T. Filmon, J. A. Birrell, O. Rüdiger, V. Fourmond, C. Léger and N. Plumeré, Reversible H2 Oxidation and Evolution by Hydrogenase Embedded in a Redox Polymer Film, *Nature catalysis*, 2021, **4**, 251–258.

52 B. E. Logan and J. M. Regan, Microbial Fuel Cells—Challenges and Applications, *Environ. Sci. Technol.*, 2006, **40**, 5172–5180.

53 I. A. Vasiliadou, A. Berná, C. Manchon, J. A. Melero, F. Martinez, A. Esteve-Nuñez and D. Puyol, Biological and Bioelectrochemical Systems for Hydrogen Production and Carbon Fixation Using Purple Phototrophic Bacteria, *Front. Energy Res.*, 2018, **6**, 8818.

54 H. Nishiyama, T. Yamada, M. Nakabayashi, Y. Maehara, M. Yamaguchi, Y. Kuromiya, Y. Nagatsuma, H. Tokudome, S. Akiyama, T. Watanabe, R. Narushima, S. Okunaka, N. Shibata, T.

Takata, T. Hisatomi and K. Domen, Photocatalytic solar hydrogen production from water on a 100-m2 scale, *Nature*, 2021, **598**, 304–307.

55  K. Villa, J. R. Galán-Mascarós, N. López and E. Palomares, Photocatalytic water splitting: advantages and challenges, *Sustainable Energy Fuels*, 2021, **5**, 4560–4569.

56  J. Na, B. Seo, J. Kim, C. W. Lee, H. Lee, Y. J. Hwang, B. K. Min, D. K. Lee, H.-S. Oh and U. Lee, General technoeconomic analysis for electrochemical coproduction coupling carbon dioxide reduction with organic oxidation, *Nature communications*, 2019, **10**, 5193.

57  S. Verma, S. Lu and P. J. A. Kenis, Co-electrolysis of $CO_2$ and glycerol as a pathway to carbon chemicals with improved technoeconomics due to low electricity consumption, *Nat Energy*, 2019, **4**, 466–474.

58  A. R. Oliveira, C. Mota, C. Mourato, R. M. Domingos, M. F. A. Santos, D. Gesto, B. Guigliarelli, T. Santos-Silva, M. J. Romão and I. A. Cardoso Pereira, Toward the Mechanistic Understanding of Enzymatic $CO_2$ Reduction, *ACS Catal.*, 2020, **10**, 3844–3856.

59  K. P. Nevin, T. L. Woodard, A. E. Franks, Z. M. Summers and D. R. Lovley, Microbial electrosynthesis: feeding microbes electricity to convert carbon dioxide and water to multicarbon extracellular organic compounds, *mBio*, 2010, **1**. DOI: 10.1128/mBio.00103-10.

60  G. Liu, F. Zheng, J. Li, G. Zeng, Y. Ye, D. M. Larson, J. Yano, E. J. Crumlin, J. W. Ager, L. Wang and F. M. Toma, Investigation and mitigation of degradation mechanisms in $Cu_2O$ photoelectrodes for $CO_2$ reduction to ethylene, *Nat Energy*, 2021, **6**, 1124–1132.

61  P. Majumdar, M. K. Bera, D. Pant and S. Patra, in *Encyclopedia of Interfacial Chemistry*, Elsevier, 2018, pp. 577–589.

62  W. Shin, S. H. Lee, J. W. Shin, S. P. Lee and Y. Kim, Highly selective electrocatalytic conversion of $CO_2$ to CO at -0.57 V (NHE) by carbon monoxide dehydrogenase from Moorella thermoacetica, *Journal of the American Chemical Society*, 2003, **125**, 14688–14689.

63  N. S. Weliwatte and S. D. Minteer, Photo-bioelectrocatalytic $CO_2$ reduction for a circular energy landscape, *Joule*, 2021, **5**, 2564–2592.

64  W. R. Leow, Y. Lum, A. Ozden, Y. Wang, D.-H. Nam, B. Chen, J. Wicks, T.-T. Zhuang, F. Li, D. Sinton and E. H. Sargent, Chloride-mediated selective electrosynthesis of ethylene and propylene oxides at high current density, *Science (New York, N.Y.)*, 2020, **368**, 1228–1233.

65  Y. Lum, J. E. Huang, Z. Wang, M. Luo, D.-H. Nam, W. R. Leow, B. Chen, J. Wicks, Y. C. Li, Y. Wang, C.-T. Dinh, J. Li, T.-T. Zhuang, F. Li, T.-K. Sham, D. Sinton and E. H. Sargent, Tuning OH binding energy enables selective electrochemical oxidation of ethylene to ethylene glycol, *Nature catalysis*, 2020, **3**, 14–22.

66  T. Burdyny and W. A. Smith, $CO_2$ reduction on gas-diffusion electrodes and why catalytic performance must be assessed at commercially-relevant conditions, *Energy Environ. Sci.*, 2019, **12**, 1442–1453.

67  D. DeSantis, B. D. James, C. Houchins, G. Saur and M. Lyubovsky, Cost of long-distance energy transmission by different carriers, *iScience*, 2021, **24**, 103495.

68  Fuell Cell Store, Water electrolysis system 5 kW. https://www.fuelcellstore.com/water-electrolysis-system-5kw, 2022.

69  A. Mayyas, M. Ruth, B. Pivovar, G. Bender and K. Wipke, NREL is a national laboratory of the U.S. Department of Energy Office of Energy Efficiency & Renewable Energy Operated by the Alliance for Sustainable Energy, LLC This report is available at no cost from the National Renewable Energy Laboratory (NREL) at www.nrel.gov/publications. Contract No. DE-AC36-08GO28308 Technical Report NREL/TP-6A20-72740 August 2019 Manufacturing Cost Analysis for Proton Exchange Membrane Water Electrolyzers. Technical Report NREL/TP-6A20-72740, 2019.

70  G. Tsotridis and A. Pilenga, EU harmonised protocols for testing of low temperature water electrolysers. JRC122565 Report., *Publications Office of the European Unio*, 2021.

71  C. H. Rivkin, R. M. Burgess and W. J. Buttner, Regulations, Codes, and Standards (RCS) for Large-Scale Hydrogen Systems. NREL/CP-5400-70929, *th International Conference on Hydrogen Safety (ICHS 2017), 11-13 September 2017, Hamburg, Germany*, 2017.

72  Z.-Z. Niu, L.-P. Chi, R. Liu, Z. Chen and M.-R. Gao, Rigorous assessment of $CO_2$ electroreduction products in a flow cell, *Energy Environ. Sci.*, 2021, **14**, 4169–4176.

73  B. Chen, B. Li, Z. Tian, W. Liu, W. Liu, W. Sun, K. Wang, L. Chen and J. Jiang, Enhancement of Mass Transfer for Facilitating Industrial-Level $CO_2$ Electroreduction on Atomic Ni–$N_4$ Sites, *Adv. Energy Mater.*, 2021, **11**, 2102152.

74  M. H. Barecka, J. W. Ager and A. A. Lapkin, Techno-economic assessment of emerging $CO_2$ electrolysis technologies, *STAR protocols*, 2021, **2**, 100889.

75  Y. C. Tan, K. B. Lee, H. Song and J. Oh, Modulating Local $CO_2$ Concentration as a General Strategy for Enhancing C−C Coupling in $CO_2$ Electroreduction, *Joule*, 2020, **4**, 1104–1120.

76  V. Rodin, J. Lindorfer, H. Böhm and L. Vieira, Assessing the potential of carbon dioxide valorisation in Europe with focus on biogenic $CO_2$, *Journal of CO2 Utilization*, 2020, **41**, 101219.

77  F. Sabatino, A. Grimm, F. Gallucci, M. van Sint Annaland, G. J. Kramer and M. Gazzani, A comparative energy and costs assessment and optimization for direct air capture technologies, *Joule*, 2021, **5**, 2047–2076.

78  A. Badgett, A. Feise and A. Star, Optimizing utilization of point source and atmospheric carbon dioxide as a feedstock in electrochemical $CO_2$ reduction, *iScience*, 2022, **25**, 104270.